\documentclass[12pt]{article}
  \usepackage{latexsym,epsfig}
\usepackage{amsfonts}
\newcommand{\Ss}{ Scherk--Schwarz }
\newcommand{\IR}{\mathbb{R}}
\newcommand{\ZZ}{\relax\ifmmode\mathchoice
{\hbox{\cmss Z\kern-.4em Z}}{\hbox{\cmss Z\kern-.4em Z}}
{\lower.9pt\hbox{\cmsss Z\kern-.4em Z}}
{\lower1.2pt\hbox{\cmsss Z\kern-.4em Z}}\else{\cmss Z\kern-.4em
Z}\fi}
\newcommand{\bfone}{\relax{\rm 1\kern-.35em 1}}
\newcommand{\cn}{\mathcal{N}}
\newcommand{\cM}{\mathcal{M}}
\else\oddsidemargin25mm\evensidemargin25mm\marginparwidth25mm\fi%
\begin{document}
\def\figurebox#1#2#3{%
        \def\arg{#3}%
        \ifx\arg\empty
        {\hfill\vbox{\hsize#2\hrule\hbox to #2{\vrule\hfill\vbox to #1{\hsize#2\vfill}\vrule}\hrule}\hfill}%
        \else
        {\hfill\epsfbox{#3}\hfill}%
        \fi}
\def\@figurecaption#1#2{\unskip\vskip10pt{#1.\hskip.5em#2\par}}
\def\thefootnote{\alph{footnote}}
\def\@makefnmark{{$^{\@thefnmark}$}}
%
%
%
%
\bibliographystyle{unsrt}    
\arraycolsep1.5pt
\def\Journal#1#2#3#4{{#1} {\bf #2}, #3 (#4)}

\def\NCA{\em Nuovo Cimento}
\def\NIM{\em Nucl. Instrum. Methods}
\def\NIMA{{\em Nucl. Instrum. Methods} A}
\def\NPB{{\em Nucl. Phys.} B}
\def\PLB{{\em Phys. Lett.}  B}
\def\PRL{\em Phys. Rev. Lett.}
\def\PRD{{\em Phys. Rev.} D}
\def\ZPC{{\em Z. Phys.} C}
\def\st{\scriptstyle}
\def\sst{\scriptscriptstyle}
\def\mco{\multicolumn}
\def\epp{\epsilon^{\prime}}
\def\vep{\varepsilon}
\def\ra{\rightarrow}
\def\ppg{\pi^+\pi^-\gamma}
\def\vp{{\bf p}}
\def\ko{K^0}
\def\kb{\bar{K^0}}
\def\al{\alpha}
\def\ab{\bar{\alpha}}
\def\be{\begin{equation}}
\def\ee{\end{equation}}
\def\bea{\begin{eqnarray}}
\def\eea{\end{eqnarray}}
\def\CPbar{\hbox{{\rm CP}\hskip-1.80em{/}}}
\arraycolsep1.5pt
%
\begin{titlepage}
\hskip 4.5cm
\vbox{}
\hskip 5.5cm
\vbox{
\hbox{CERN-TH/2003-163}
\hbox{SPIN-03/24}
\hbox{ITP-UU-03/38}
}
\vskip 3cm
\begin{center} {\Large \bf Fluxes, supersymmetry breaking and gauged supergravity  }
\vskip 3cm {\bf Laura Andrianopoli$^\diamondsuit$, Sergio Ferrara$^\clubsuit$ and Mario Trigiante$^\spadesuit$ }\\
\skip 0.5cm
{ \it $^\diamondsuit$ CERN, Theory Division, CH-1211 Geneva 23, Switzerland,} \\
{E-mail: Laura.Andrianopoli@cern.ch} \\
 {\it $^\clubsuit$ 
CERN, Theory Division, CH-1211 Geneva 23, Switzerland and}\\
{\it  Laboratori Nazionali di Frascati, INFN, Italy,}\\
{E-mail: Sergio.Ferrara@cern.ch} \\
 {\it $^\spadesuit$ Spinoza Institute, Leuvenlaan 4 NL-3508, Utrecht, The Netherlands,}\\
{E-mail: M.Trigiante@phys.uu.nl}

\vskip 2cm

\begin{abstract}
\noindent
We report on the gauged supergravity interpretation of certain compactifications of superstring theories 
with p-form fluxes turned on.\\ \noindent
We discuss in particular the interplay of duality symmetries in type IIB orientifolds and gauged isometries in
the corresponding supergravity models.
\\
\noindent
Turning on fluxes is generally described by the gauging of some nilpotent Lie group whose generators correspond to axion 
symmetries of R-R and N-S scalars.
\end{abstract}

\vskip 2cm 
Contribution to the proceedings of ``Sugra20'' Conference, Department of Physics, 
Northeastern University, Boston (Ma) 02115 USA; March 2003
\end{center}
\end{titlepage}

\vfill\eject

\section{Introduction}
Gauged supergravities in four dimensions have been the subject of a renewed interest in recent time, 
especially because of their connection with higher dimensional theories compactified on 
manifolds which allow fluxes of p-forms along the internal directions.

A particularly appealing class of such theories is given by type IIA and IIB compactifications on 
 Ricci-flat manifolds which, in absence of fluxes, leave $N=8,4,2$ unbroken supersymmetries in four dimensions.
Turning on fluxes develops a scalar potential in these theories, which usually gives moduli stabilization, reduced supersymmetry
and, in certain cases, leaves a vanishing cosmological constant \cite{ps}--\cite{lm}.
In fact, theories with vanishing cosmological constant are generalized no-scale models, which were studied long ago in the pure supergravity framework
\cite{cfkn,elnt}.

Another interesting class of models  recently studied in the literature is given by
compactifications on $T_6$ and $K3 \times T_2$ of IIB orientifolds with three-form fluxes turned on \cite{gkp,t6z2red,tt}.
These theories have $N=4$ and $N=2$ unbroken supersymmetries respectively, and the presence of fluxes may give rise to partial supersymmetry breaking 
$N=4,2 \to N=1,0$ with vanishing vacuum energy.

Due to the precise knowledge of the moduli space of these theories, an exact estimate of the (tree-level) scalar potential is possible in all these models.
 The flux-compactifications share the  feature that their low energy description is given in terms of an 
extended supergravity with mass deformations and a scalar potential 
\cite{t6z2,porrati,quaternions,k3t2}.

In the supergravity framework this can be achieved through the gauging of some of the isometries of the non linear $\sigma$-model spanned by the scalar sector.
Indeed, the relevant isometries which are gauged in presence of fluxes are those associated to axion symmetries of those scalars 
coming from R-R forms $C^{(p)}$ ($p=0,2,4$) and from the N-S two-form $B$ field.
In fact the latter only contributes in the generalized case of orientifolds of $T_6$ where the orientifold projection 
acts only on some directions of the $T_6$ torus \cite{ourlast}.

A similar phenomenon occurs in the Scherk--Schwarz generalized compactification of M-theory \cite{ss}.
Indeed, \Ss spontaneously broken supergravity can be shown to be completely equivalent to a gauged supergravity with a non semisimple gauge group
given by the semi-direct product of a $U(1)$ factor, gauged by the Kaluza--Klein vector, times axionic symmetries corresponding to 
isometries in the scalar sector coming from five dimensional vectors \cite{Andrianopoli:2002mf,deWit:2002vt,Dabholkar:2002sy} (see \cite{deWit:2002vt} also for the construction of the five--dimensional gauged supergravity which describes the spontaneously broken model deriving from $D=6\,\rightarrow\, D=5$ \Ss dimensional reduction).
The $U(1)$ symmetry is related to the central charge of the  BPS massive representations of the spontaneously broken theory \cite{Andrianopoli:2002mf,deWit:2002vt,Dabholkar:2002sy}.

In all the above mentioned compactifications, the underlying supergravity relies on a gauge group which is a certain 
subgroup of the duality group having a linear (adjoint) action on the vector fields of the theory.
Since the most general duality transformation in $D=4$ is an element of $Sp(2n_V,\IR)$ 
(where $n_V$ denotes the number of vectors in the theory), 
the gauge group must be a {\em lower block-triangular} symplectic matrix, in order not to mix electric with magnetic fields.

The same ungauged $N$-extended supergravity can therefore have several inequivalent deformations, corresponding to 
different choices of the gauge group.
The latter is chosen by selecting, among the duality symmetries, the ones which are realized linearly on the vector fields of the theory.

\section{Duality rotations and gauging}
According to Gaillard and Zumino \cite{gz}, the most general linear transformation between electric and magnetic field strengths $F^\Lambda, G_\Lambda$ ($\Lambda =1, \cdots , n_V$), in a given four dimensional theory with vectors coupled to scalar fields,
is a $Sp(2n_V, \IR)$ transformation, acting on the self-dual components $F^{+\Lambda}, G_{\Lambda}^+$ \footnote{where a generic
 self-dual field strength $T^+$ is defined by
$ T^+_{\mu\nu} \equiv \frac 12 (T_{\mu\nu} +\frac {\rm i}2 \epsilon_{\mu\nu\rho\sigma}T^{\rho\sigma})$
}
as
\be
\pmatrix{
F^\Lambda\cr G_\Lambda}^\prime=
\pmatrix{
A^\Lambda_{\ \Sigma}&B^{\Lambda\Sigma}\cr C_{\Lambda\Sigma}& D
_\Lambda^{\ \Sigma}}\pmatrix{
F^{\Sigma}\cr G_{\Sigma}},\label{dual}\ee
where the matrix $S=\pmatrix{A&B\cr C&D}$ is symplectic \footnote{that is $S^T \Omega S =\Omega$, which implies, by choosing as symplectic metric $\Omega = \pmatrix{
0 & \bfone \cr -\bfone &0}$, that $(A^TC)$, $(B^TD)$ are symmetric and $A^TD-C^TB =\bfone$.
Passing to the algebra, for $S=\bfone +s$, $s = \pmatrix{ a&b\cr c&d}$, this implies $b^T=b$, $c^T =c$, $d^T =-a$.}.

If we define a complex symmetric matrix $\cn_{\Lambda\Sigma}$ through the equation
\be G^+_\Lambda = \cn_{\Lambda\Sigma} F^{+\Sigma}\ee
then under a duality rotation (\ref{dual}) $\cn$ gets transformed to
\be \cn^\prime = (C+D\cn) \cdot (A+B\cn)^{-1}. \ee

The matrix $\cn$ appears in the vector Lagrangian as 
\be \mathcal{L}=2\Im (\cn_{\Lambda\Sigma} F^{+\Lambda} F^{+\Sigma})= 2\Im (G^+_\Lambda F^{+\Lambda}).\ee
For infinitesimal transformations we have
\bea
\delta F^{+\Lambda} &=& a^\Lambda_{\ \Sigma}F^{+\Sigma} + b^{\Lambda\Sigma} G^+_\Sigma, \qquad b^{\Lambda\Sigma}= b^{\Sigma\Lambda};
\nonumber\\
\delta G^{+}_{\Lambda} &=& c_{\Lambda \Sigma}F^{+\Sigma} -a^{\Sigma}_{\ \Lambda} G^+_\Sigma,
 \qquad c_{\Lambda\Sigma}= c_{\Sigma\Lambda};\nonumber\\
\delta \cn &=& c-a^T \cn - \cn a - \cn b \cn .
\eea
The most general subgroup of $Sp(2n_V, \IR)$ leaving the Lagrangian invariant up to a total derivative is obtained by setting $B=b=0$.
In this case the vector fields $A^\Lambda$ (such that $F^\Lambda =dA^\Lambda$) transform as
\be \delta A^\Lambda = a^\Lambda_{\ \Sigma}A^{\Sigma}\ee
and
\be \delta \cn = c-a^T \cn - \cn a .
\ee
A duality transformation can be chosen as gauge symmetry if it can be found as a certain subalgebra of those duality symmetries 
which leave the action invariant, that is only if $B=0$.

Axion symmetries have $C\neq 0$. In this case the gauge group can be non-abelian or abelian  depending on whether the block $a$ is different from
 zero or not. Moreover, axion isometries are usually embedded in nilpotent groups, and this implies that the matrix $a$ itself is also lower (or upper) triangular.

$\sigma$-models coupled to vectors have a richer structure than ordinary $\sigma$-models.
For coset spaces $G/H$, $G$ acts on the vector field strengths and their duals in a symplectic representation. 
In particular, in this representation the coset representatives are symplectic matrices $L(\phi) = \pmatrix{A(\phi)& B(\phi)\cr
C(\phi)&D (\phi)}$  through which, by defining
\be f(\phi) = \frac 1{\sqrt 2}(A-{\rm i}B) \, , \qquad h(\phi) = \frac 1{\sqrt 2}(C-{\rm i}D), \ee
one gets $\cn = h \cdot f^{-1}$ as an explicit, scalar dependent, expression for the $\cn$ matrix.
This formula allows the computation of the vector Lagrangian from first principles.

A simplification here occurs by choosing a solvable parametrization of the coset in the symplectic representation.
In this case $B(\phi) =0$ and  $D(\phi)=(A^T)^{-1}$, so that
\be \cn = \left[C-{\rm i} (A^{ T})^{-1}\right] \cdot A^{-1} = C\cdot A^{-1} - {\rm i} (A\cdot A^T)^{-1} .\ee

Let us now suppose to gauge some subalgebra of the $(a,c)$ generators which has an adjoint action on the vectors (just trivial in the
abelian case).
Then, in such case
\be a^\Lambda_{\ \Sigma} = f^\Lambda_{\ \Sigma\Gamma} \xi^\Gamma \, ; 
\quad c_{\Lambda\Sigma} = c_{\Lambda\Sigma,\Gamma}\xi^\Gamma 
\ee
where the constants $c_{\Lambda\Sigma,\Gamma}$ satisfy $c_{(\Lambda\Sigma,\Gamma)}=0$ and an additional cocycle condition in the non abelian case, where $f^\Lambda_{\ \Sigma\Gamma} \neq 0$:
\begin{eqnarray}
\frac{1}{2}\,f^\Gamma_{\Lambda\Sigma}\, c_{\Delta\Omega,\Gamma}+c_{\Gamma\Omega,[\Sigma} f^\Gamma_{\Lambda]\Delta}+c_{\Gamma\Delta,[\Sigma} f^\Gamma_{\Lambda]\Omega}&=&0
\end{eqnarray}
 These conditions appear because of the fact that, when $c$ generators (axions) are gauged,
a gauge- and supersymmetry-invariant Lagrangian can be found only after adding to the Lagrangian a Chern-Simons type term, proportional to 
$c_{\Lambda\Sigma,\Gamma}$ \cite{dlv}.

The above is the general structure of the gauge transformations in extended supergravity when the axion symmetries are gauged.

This is a feature common to all orientifold models with fluxes turned on as well as to models obtained from \Ss generalized dimensional
 reduction.

\section{$T_6$ orientifolds, nilpotent algebras and gaugings}
In the present section we shall briefly review  the construction, which was accomplished in \cite{ourlast}, of some new four dimensional orientifold models (both in type IIB and IIA) where the orientifold projection involves an orbifold projection with respect to the space--inversion $I_{9-p}$ on $9-p$ coordinates, transverse to the $Dp$-brane world volume. Since in this construction the $Dp$--brane world volume fills the non-compact space-time, the torus $T_6$ is naturally split into Neumann directions $T_{p-3}$ and Dirichlet directions $T_{9-p}$. These models therefore will be denoted by $T_{p-3}\times T_{9-p}$. Their low-energy descriptions (in absence of fluxes and $D$--branes) are all given in terms of a four 
dimensional ungauged ${\mathcal N}=4$ supergravity coupled to six vector multiplets from the closed string sector. In the following discussion we shall always restrict ourselves to the bulk degrees of freedom. The scalar fields span a manifold with the following $G/H$ form:
\begin{eqnarray}
{\mathcal M}_{{\mathcal N}=4}&=&\frac{SL(2,\mathbb{R})}{SO(2)}\times \frac{SO(6,6)}{SO(6)\times SO(6)},
\end{eqnarray}
the duality group being $G=SL(2,\mathbb{R})\times SO(6,6)$. We summarize in Table 1 the bosonic field content of these models in the type IIB and IIA settings respectively (in our conventions the directions of $T_6$ are labeled by an index $n=1,\dots, 6$ whereas the directions of $T_{p-3}$ and of $T_{9-p}$ are labeled respectively by indices $i=1,\dots , p-3$ and $a=p-2,\dots , 9$ and the non--compact  space--time directions by Greek letters).
\begin{table}[ht]
\vskip 0.5 cm
\caption{Massless degrees of freedom for the IIB/IIA 
orientifolds}
\begin{center}
\begin{tabular}{|r|c|c|c|}
\hline
$p$ & scalars & vectors & type \\
\hline\hline
9 & $g_{ij}$, $\phi$, $C_{\mu\nu}$, $C_{ij}$ & $G^{i}_{\mu}$, $C_{i\mu }$& \\
\cline{1-3}
7 & $g_{ij}$, $g_{ab}$, $\phi$, $B_{ia}$, $C$, $C_{ia}$, $C_{ijkl}$, $C_{ijab}$
& $G^{i}_{\mu}$, $B_{a\mu }$, $C_{a\mu}$, $C_{ijk\mu}$& IIB\\
\cline{1-3}
5 & $g_{ij}$, $g_{ab}$, $\phi$, $B_{ia}$, $C_{\mu\nu}$, $C_{ij}$, $C_{ab}$,
$C_{iabc}$ & $G^{i}_{\mu}$, $B_{a\mu}$, $C_{i\mu }$, $C_{abc\mu}$& \\
\cline{1-3}
3 & $g_{ab}$, $\phi$, $C$, $C_{abcd}$ & $B_{a\mu}$, $C_{a\mu }$& \\
\hline\hline
8 & $g_{ij}$, $g_{66}$, $\phi$, $B_{i6}$, $C_{i}$, $C_{6\mu\nu}$, $C_{ij6}$ & 
$G^{i}_{\mu}$, $C_{\mu}$, $C_{i6\mu}$, $B_{6\mu}$& 
\\
\cline{1-3}
6 & $g_{ij}$, $g_{ab}$, $\phi$, $B_{ia}$, $C_{a}$, $C_{i\mu\nu}$, $C_{ijk}$,
$C_{iab}$
& $G^{i}_{\mu}$, $B_{a\mu }$, $C_{ij\mu}$, $C_{ab\mu}$& IIA\\
\cline{1-3}
4 & $g_{11}$, $g_{ab}$, $\phi$, $B_{1a}$, $C_{1}$, $C_{a\mu\nu}$, $C_{abc}$
& $G^{1}_{\mu}$, $B_{a\mu}$, $C_{\mu }$, $C_{1a\mu}$ &\\
\hline
\end{tabular}
\end{center}
\end{table}
The fluxes allowed by the orientifold projections are listed in Table 2.
\begin{table}[ht]
\vskip 0.5 cm
\caption{Allowed fluxes for the IIB/IIA 
orientifolds. $F$, $H$ and $G$ fluxes are associated to the $B$, $C_2$ and 
$C_4$ fields}
\vskip 0.5 cm
\begin{center}
\begin{tabular}{|r|c|c|}
\hline
$p$ & fluxes &  type\\
\hline\hline
9 & none & \\
\cline{1-2}
7 & $H_{aij}$, $F_{aij}$, $G_{abijk}$ &
 IIB\\
\cline{1-2}
5 & $H_{abc}$, $F_{iab}$, $H_{ija}$, $G_{ijabc}$
& \\
\cline{1-2}
3 & $H_{abc}$, $F_{abc}$ &
\\\hline\hline 8 & $H_{ij6}$, $G_{ijk6}$
&\\
\cline{1-2}
6 & $H_{ija}$, $H_{abc}$, $F_{ia}$, $G_{ijab}$
& IIA\\
\cline{1-2}
4 & $H_{abc}$, $F_{ab}$, $G_{1abc}$ &\\ \hline
\end{tabular}
\end{center}
\end{table}

The  ${\mathcal N}=4$ orientifold models (in the absence of fluxes) can be consistently
constructed as truncations of the unique four dimensional
${\mathcal N} =8$ supergravity which describes the low-energy
limit of toroidally compactified type II superstrings. In the scalar-field sector 
this amounts to defining the embedding of the ${\mathcal N} =4$ scalar manifold into the ${\mathcal N} =8$ one. The duality
 group of the latter theory is $E_{7(7)}$ which 
acts non-linearly on the $70$ scalar
fields, and linearly, as a $Sp(56,\mathbb{R})$ symplectic
transformation, on the  $28$ electric field strengths and their
magnetic dual. Within this theory an intrinsic group-theoretical
characterization of the ten dimensional origin of the scalar and
vector fields can be achieved. In the so-called solvable Lie
algebra representation of the scalar sector
\cite{Andrianopoli:1996bq,Cremmer:1997ct,Bertolini:1999uz}, the scalar
manifold
\begin{equation}
{\mathcal M}_{\rm {\mathcal N} =8} = \exp{(Solv(e_{7(7)}))},
\end{equation}
is expressed as the group manifold generated
by the solvable Lie algebra $Solv(e_{7(7)})$ defined by the Iwasawa
decomposition of the $e_{7(7)}$ algebra:
\begin{equation}
e_{7(7)} = su(8)+Solv(e_{7(7)}) \,.
\end{equation}
In this framework, there is a natural one-to-one correspondence
between the scalar fields and the generators of $Solv(e_{7(7)})$.
The latter consist of the $e_{7(7)}$ Cartan generators $H_p$, parametrized by the $T_6$ radii $R_n
=e^{\sigma_n}$ together with the dilaton $\phi$, and of the shift
generators corresponding to the $63$  positive roots $\alpha$ of
$e_{7(7)}$, which are in one-to-one correspondence with the
axionic scalars that parametrize them. This correspondence between
Cartan generators and positive roots on one side and scalar fields on the
other, can be pinpointed by decomposing $Solv(e_{7(7)})$ with
respect to some relevant groups \cite{Andrianopoli:1996bq}. 
The axions deriving from
ten--dimensional tensor fields (i.e. the Kalb--Ramond form
$B_{MN}$ and the R--R fields) transform in tensorial
representations of the $SL(6,\mathbb{R})_g$ isometry group of the
$T_6$ metric moduli and so do the corresponding solvable
generators. If we express the  $e_{7(7)}$ roots
with respect to an orthonormal basis of Euclidean vectors
$\{\epsilon_r\}$, $r=1,\dots ,7$,
 the precise correspondence between axions and
$e_{7(7)}$ nilpotent generators reads:
\begin{eqnarray} C_{n_1n_2\dots n_k}
&\leftrightarrow & T^{n_1n_2\dots n_k}\,=\,E_{a +\epsilon_{n_1}+\dots \epsilon_{n_k}} \,,
\nonumber
\\
C_{ n_1n_2\dots n_k\mu\nu} &\leftrightarrow &  T^{m_1m_2\dots m_{6-k}}\,=\,E_{a
+\epsilon_{m_1}+\dots
\epsilon_{m_{6-k}}}\,,\,\,\,\,\,\,(\epsilon^{n_1\dots n_k m_1\dots
m_{6-k}} \neq 0) \,,\nonumber
\\
B_{n m} &\leftrightarrow & T_B^{n m}\,=\,E_{ \epsilon_{n}+\epsilon_{m}} \,,\nonumber
\\
B_{\mu\nu} &\leftrightarrow & T \,=\,E_{\sqrt{2}\,\epsilon_7 }\,, \nonumber
\\
G_{n m}&\leftrightarrow
&T^n{}_m \,=\,E_{\epsilon_{n}-\epsilon_{m}}\,,\,\,\,\,\,\,(n\neq m) \,,\nonumber
\end{eqnarray}
where
\begin{equation}
a = -{\textstyle\frac{1}{2}} \sum_{i=1}^6\,
\epsilon_i+{\textstyle\frac{1}{\sqrt{2}}}\,\epsilon_7 \,.
\end{equation}
The torus radii $R_n=e^{\sigma_n}$ and the ten--dimensional dilaton $\phi$ 
enter this description in the following combination with the Cartan generators:
\begin{eqnarray}
\vec{h}\cdot \vec{H}&=&\sum_{p=1}^7\,h^p\,H_{\epsilon_p}\,=\, \sum_{n=1}^6\,\sigma_n\,H_{(\epsilon_n+\frac{1}{\sqrt{2}}\epsilon_7)}-\frac{\phi}{2}\, H_a.
\end{eqnarray}
so that the kinetic term of the axion corresponding to the root $\alpha$ contains the correct exponential factor $e^{-2\vec{h}\cdot \vec{\alpha}}$.\par As far as the
scalar sector is concerned, the embedding of the ${\mathcal N}=4$
orientifold models $T_{p-3}\times T_{9-p}$ (in absence of fluxes)
inside the ${\mathcal N}=8$ theory (in its type IIA or IIB
versions) is defined by the condition that the embedding of the ${\mathcal
N}=4$ duality group $SL(2,\mathbb{R})\times SO(6,6)$ inside the
$E_{7(7)}$  fulfill the following condition:
\begin{equation}
SO(6,6)\,\cap\,  GL(6,\mathbb{R})_g = O(1,1)\times
SL(p-3,\mathbb{R}) \times  SL(9-p,\mathbb{R}) \,. \label{inter}
\end{equation}
\begin{center}
\begin{figure}[t]
\epsfxsize=30pc 
\epsfbox{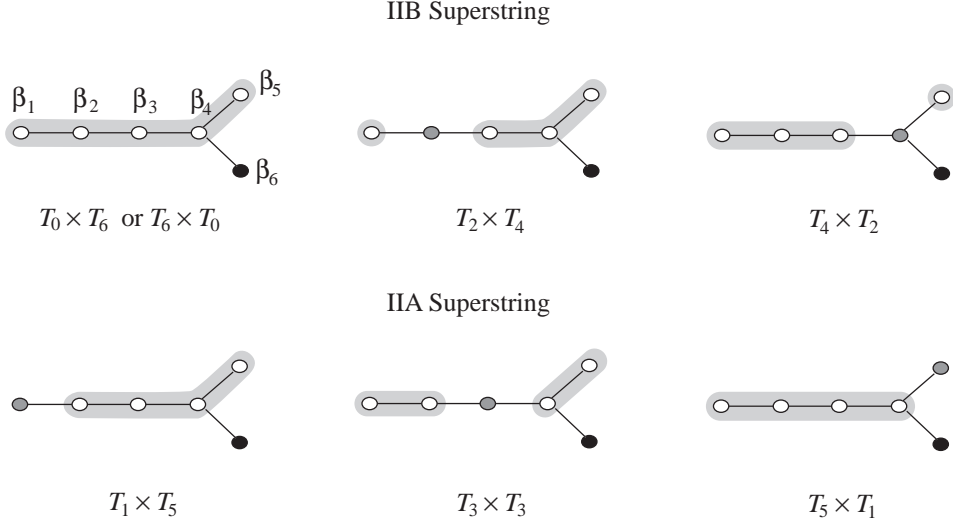}
\caption{$SO(6,6)$ Dynkin diagrams for the $T_{p-3}\times T_{9-p}$
models. The shaded sub-diagrams define the groups $SL(p-3,
\mathbb{R})\times SL(9-p,\mathbb{R})$ acting transitively on the
metric moduli. The empty circles define simple roots corresponding
to the metric moduli $g_{ij},\,g_{ab}$, the gray circle denotes a
simple root corresponding to a Kalb--Ramond field $B_{ia}$ and the
black circle corresponds to a R--R axion. \label{o66}}
\end{figure}
\end{center}
 Condition (\ref{inter}) amounts to requiring that the $T_6$ metric moduli in our models are related either to  the $T_{p-3}$ metric $g_{ij}$ or to the $T_{9-p}$ metric $g_{ab}$. It fixes the 
ten--dimensional interpretation of the fields  in the ungauged ${\mathcal N}=4$ models (except for the cases $p=3$ and $p=9$), and for a given $p$ it is 
consistent with the bosonic spectrum resulting from the corresponding 
orientifold reductions. In the $p=3$ and $p=9$ cases, the two
embeddings are characterized by a different interpretation of the 
scalar fields, related to the $T_6/\mathbb{Z}_2$ orientifold reduction 
in the presence of D3 or D9 branes. The axions not deriving from the internal metric consist of $ C_{i_1\dots i_{p-3}}$ in the 
external $SL(2,\mathbb{R})/SO(2)$ factor, $(p-3)\,(9-p)$ moduli $B_{i a}$ in 
the bi-fundamental of $SL(p-3,\mathbb{R})\times  SL(9-p,\mathbb{R})$ and $15$ 
R--R moduli which we shall generically denote by $C_I$ and which span
the maximal abelian ideal $\{T^I\}$ of $Solv(so(6,6))$. The scalars $B_{ia}$ 
and $C_I$ parametrize a $15+(p-3)\,(9-p)$ dimensional subalgebra $N_p$ of 
$Solv(so(6,6))$ consisting of nilpotent generators only.\par
Switching on fluxes amounts in the four--dimensional theory to introducing a suitable gauge group contained inside $N_p$. Let us first list for the various models the generator content of $N_p$ and their algebraic structure.

\noindent {{\em $T_0\times T_6$ model}} --- In this case $N_3$ is abelian and its generic element is  $C_{abcd}\, T^{abcd}$.

\noindent {{\em $T_2\times T_4$ model}} --- The algebra $N_5$ is 23--dimensional and has the following structure:
\begin{eqnarray}
N_5&\equiv &\{C_{\mu\nu}\,T+B_{ia}\, T_B^{ia}+C_{iabc}\, T^{iabc}+C_{ab}\, T^{ab}\}\,,\nonumber\\
&&\left[T_B^{ia},\, T^{bc}\right]\,=\, T^{iabc}\,,\,\,\,\,\,\left[T^{ia}_B,\, T^{jbcd}\right]\,=\,\epsilon^{ij}\, \epsilon^{abcd}\, T.
\end{eqnarray}

\noindent {{\em $T_4\times T_2$ model}} --- The algebra $N_7$ is 23--dimensional and has the following structure:
\begin{eqnarray}
N_7&\equiv &\{C_{(0)}\,T_0+B_{ia}\, T^{ia}_B+C_{ia}\, T^{ia}+C_{ijab}\, T^{ijab}\}\,,\nonumber\\
&&\left[T_0,\, T^{ia}_B\right]\,=\, T^{ia}\,,\,\,\,\,\left[T^{ia}_B,\, T^{jb}\right]\,=\,T^{ijab}.
\end{eqnarray}

\noindent {{\em $T_6\times T_0$ model}} --- In this case $N_9$ is abelian and its generic element is  $C_{ab}\, T^{ab}$.

\noindent {{\em $T_1\times T_5$ model}} --- The algebra $N_4$ is 20--dimensional and has the following structure:
\begin{eqnarray}
N_4&\equiv &\{C_{a\mu\nu }\,T_a + B_{1a}\, T^{a}_B+C_{abc}\, T^{abc}\}\,,\nonumber\\
&&\left[T^{abc},\, T^{d}_B\right]\,=\, \epsilon^{abcde}T_{e}.
\end{eqnarray}

\noindent {{\em $T_3\times T_3$ model}} --- The algebra $N_6$ is 24--dimensional and has the following structure:
\begin{eqnarray}
N_6&\equiv &\{C_{ a}\,T^a + B_{ia}\, T^{ia}_B+C_{iab}\, T^{iab}+C_{i\mu\nu}\, T_{i}\}\,,\nonumber\\
&&\left[T^{a},\, T^{ib}_B\right]\,=\, T^{iab}\,,\,\,\,\,\left[T_B^{ia},\, T^{jbc}\right]\,=\,\epsilon^{ijk}\,\epsilon^{abc}\, T_{k}.
\end{eqnarray}

\noindent {{\em $T_5\times T_1$ model}} --- The algebra $N_8$ is 20--dimensional and has the following structure:
\begin{eqnarray}
N_6&\equiv &\{C_{ i}\,T^i + B_{i6}\, T^{i6}_B+C_{ij6}\, T^{ij6}\}\,,\nonumber\\
&&\left[T^{i},\, T^{j6}_B\right]\,=\, T^{ij6}.
\end{eqnarray}
\vskip 0.5 cm
\noindent {{\em Fluxes and gaugings: a preliminary analysis}} --- Let us consider the $T_2\times T_4$ model in presence of the fluxes $H_{ija}=\epsilon_{ij}\, H_a$ and $F_{iab}$. These fluxes appear as structure constants 
\begin{eqnarray}
\left[X_i,\,X_j\right] &=& \epsilon_{ij}\, H_a\, X^a\,,
\nonumber \\
\left[X_i,\,X^a\right] &=& F_i^{ab}\, X_b \,, \label{algebra2}
\end{eqnarray}
of a gauge algebra ${\mathcal G}_g\,\equiv\,\{X_i,\,X^a,\,X_a\}$ with connection
$\Omega^g_\mu = G^i_\mu \, X_i+B_{a\mu}\, X^a+ C^a_\mu\,X_a$, all other
commutators vanishing.

The identification
\begin{eqnarray}
\label{gauge2} X^\prime_i &=&-F_i^{ab}\, T_{ab}+H_a\, T_B{}^a_{i} \,,
\nonumber \\
X^{a\prime } &=& F_i^{ab}\, T^i_b \,,
\end{eqnarray}
of the gauge generators with the isometries of the solvable algebra  (in our conventions $F_{i}^{ab} = \frac{1}{2} 
\epsilon^{abcd}\,F_{i cd}$ and $T^i_a=(1/3!)\,\epsilon_{abcd}\,T^{ibcd} $), 
reproduces only a contracted version of the algebra (\ref{algebra2}) in which three of the central charges $X_a$ vanish and we are left with $X^\prime_a\,=\,- H_a\, T$. If we denote by $\{X^\perp\}=\, \{X_a\}/ \{X^\prime_a\}$ these three central generators, we see that the subgroup ${ \mathcal G}^\prime_g\,=\,\{X^\prime_i,\,X^{a\prime },\,X^\prime_a\}$ of the isometry group
 which is gauged coincides with the quotient:
\begin{equation}
 {\mathcal G}^\prime_g\,= \,{\mathcal G}_g/\{X^\perp\} \,,
\end{equation}
that amounts to imposing the vanishing of the central terms $X^\perp$ on all fields.
One can verify that the vectors
$G^i_\mu$, $B_{a\mu}$  and $C^a_\mu$ transform in the co-adjoint
representation of ${\mathcal G}_g$. The non-abelian field strengths are:
\begin{eqnarray}
{\mathcal H}_{a\mu\nu} &=& \partial_\mu B_{a\nu}-\partial_\nu B_{a\mu}-
\epsilon_{ij}\,H_a\, G^i_\mu\,G^j_\nu \,,
\nonumber\\
F^a_{\mu\nu}&=& \partial_\mu C^a_{\nu}-\partial_\nu C^a_{\mu}+F_i^{ab}\, 
G^i_\mu\,B_{b\nu}-F_i^{ab}\, G^i_\nu\,B_{b\mu} \,,
\nonumber\\
{\mathcal F}^i_{\mu\nu}&=&\partial_\mu G^i_{\nu}-\partial_\nu G^i_{\mu} \,.
\end{eqnarray}
and the covariant derivatives for the scalar fields read:
\begin{eqnarray}
D_\mu c&=&\partial_\mu c+H_a\,
C^a_\mu -B_{a\mu}\,
F_i^{ab}\,B_{jb}\,\epsilon^{ij} \,,
\nonumber\\
D_\mu C^a_i&=&\partial_\mu C^a_i-B_{b\mu}\, 
F_i^{ba}-G^j_\mu\,F_j^{ab}\, B_{bi} \,,
\nonumber\\
D_\mu B_{ia}&=&\partial_\mu B_{ia}+G_{i\mu}\,H_a\,,\nonumber\\
D_\mu C_{ab}&=&\partial_\mu C_{ab}+G_\mu^i\, F_{iab} \,.
\end{eqnarray}
As far a the $T_4\times T_2$ model is concerned let us consider the gauging 
which corresponds to turning on the fluxes $F_{ija},\, H_{ija}$. It is useful 
to describe 
the $H$ and $F$ forms as elements of an $SL(2,\mathbb{R})$ doublet labeled by an index $\alpha=1,2$, the couple of indices $(\alpha,\,a)$ span the representation ${\bf 4}$ of $SO(2,2)$. We shall then denote the flux--forms by $H^\lambda_{ij}$ where $\lambda $ is the index of the ${\bf 4}$ in which the invariant metric is diagonal $\eta_{\lambda\lambda^\prime}\equiv {\rm diag}(-1,-1,+1,+1)$. Similarly the generators $T_B^{ia},\, T^{ja}$ will be denoted by $T^{\lambda i}$ and the vectors $B_{a\mu},\,C_{a\mu}$ by $A^\lambda_\mu$. Inspection of the dimensionally reduced three--form kinetic term indicates 
for the four--dimensional theory a gauge group $G$ with connection $\Omega=X_i\, G^i_\mu+X_\lambda A^\lambda_\mu$ and the following structure:
\begin{eqnarray}
\left[X_i,\,X_j\right]&=& H_{ij}^\lambda X_\lambda.
\label{galg}\end{eqnarray}
If we identify the gauge generators with isometries as follows:\begin{eqnarray}
X_{i} &=& -H^{\lambda}_{ij}\,T^j_\lambda\,,
\nonumber \\
X^\lambda &=& {\textstyle\frac{1}{2}} \,H^{\lambda^\prime }_{ij}\,T^{ij} \,,
\end{eqnarray}
 then it can be shown that they close the algebra (\ref{galg}) only if $H_{ij}^\lambda\,H^{ij}_\lambda\,=\,0$ which amounts to requiring that $\int_{T_6}\, H_{(3)}\wedge F_{(3)}\,=\,0$ (this condition is consistent with a constraint found in \cite{toappear} on the embedding matrix of a new gauge group in the ${\mathcal N}\,=\,8$ theory, which seems to yield an ${\mathcal N}\,=\, 8$ ``lifting'' of the type IIB orientifold models $T_{p-3}\times T_{9-p}$ discussed here). The vectors transform indeed in the co--adjoint of $G$ and the non--abelian field strengths are:
\begin{eqnarray}
F^\lambda_{\mu\nu} & = &\partial_\mu A^\lambda_\nu-\partial_\nu
A^\lambda_\mu -H_{ij}^\lambda\,G^i_\mu\,G^j_\nu \,,
\nonumber\\
{\mathcal F}^i_{\mu\nu} &=&\partial_\mu G^i_\nu-\partial_\nu G^i_\mu \,,
\nonumber\\
F^i_{\mu\nu} &=& \epsilon^{ijkl}\,\left(\partial_\mu
C_{jkl\nu}-\partial_\nu C_{jkl\mu}\right) \,.
\end{eqnarray}
The relevant covariant derivatives for the axions read:
\begin{eqnarray}
D_\mu C_{ij} &=& \partial_\mu
C_{ij}-{\textstyle\frac{1}{2}}\,H_{ij\,\lambda}\,A^{{\lambda}}_{\mu}+\frac{1}{2}\,G_\mu^k\,
H^{\lambda}_{k[i}\,
\Phi_{j]\lambda}\,,
\nonumber\\
D_\mu\Phi^\lambda_i &=& \partial_\mu \Phi^\lambda_i
-H^\lambda_{ij}\,G_\mu^j \,.
\end{eqnarray}

\section{Supersymmetry breaking in $T_6 / Z_2$ and $K3 \times T_2 /Z_2$ orientifolds}
In this section we are going to describe the basic features of 
the supersymmetry breaking pattern induced by the presence of non trivial 
3-form fluxes in two orientifold compactifications of IIB superstring.
In particular, we will focus on the bulk sector of the theories which 
is the one 
responsible for the supersymmetry breaking and moduli stabilization, without considering here the D-brane degrees of freedom.

Before entering into the details, let us briefly summarize the conditions for the vacuum to preserve some unbroken supersymmetry.
They are exhausted by the request that, in the vacuum, the supersymmetry transformation laws of all the fermions $f$ are zero for
the corresponding supersymmetry parameter $\epsilon$, that is
\be
 <\delta f >=0
\Rightarrow \left\{\matrix{ \delta \psi_A = 0 \; : \quad <S_{AB}(\phi,g)>\epsilon^B = - \sqrt{\frac { <V(\phi) >}{6} }\
\epsilon_A
\cr \delta \lambda^{IA} =0 \; : \quad <N^{IA}(\phi,g)>\epsilon_A=0}\right. 
\ee
where $\psi_A$ are the gravitini and  $\lambda^{IA}$  generic spin 1/2 fields,while $S_{AB}$ and $N^{IA}$ denote
 the gravitino mass matrix and fermionic shifts respectively.

 $V(\phi)$ is the scalar potential defined by
$$\delta^A_B V(\phi)=-3 \bar S^{AC}S_{BC}+N^{IA} N_{IB}$$ 

Both the models considered in this section 
 have a positive definite scalar potential, allowing vacuum configurations 
with partial super Higgs and zero vacuum energy. This is related to the fact that, in both cases, 
the gauging responsible for the presence of a scalar potential corresponds to switching on charges 
only for translational isometries of the scalar manifolds. It only depends, in fact, on universal properties leading to
cancellation of positive and negative contributions in the scalar potential as it occurred in $N=1$ no-scale models
\cite{cfkn,elnt}

\subsection{The $T_6 /Z_2$ orientifold}
The bulk sector  of the $T_6 /Z_2$ orientifold model is described by $N=4$ supergravity coupled to 6 vector multiplets, where,
as discussed in section 3,  the 38 scalars parametrize the coset manifold $SL(2,\IR) /U(1) \times SO(6,6)/[SO(6)\times SO(6)]$ (for a derivation of this model from dimensional reduction of type IIB theory see \cite{t6z2red} and references therein).

Its gauged supergravity description (first constructed in \cite{t6z2}) relies on an abelian gauge algebra
which corresponds to a 12 dimensional subalgebra of the 15 shift symmetries of the scalars coming from the R-R 4-form  
$C_{\Gamma\Delta\Pi\Omega} \equiv \epsilon_{\Lambda\Sigma\Gamma\Delta\Pi\Omega}B^{\Lambda\Sigma}$; $\Lambda , \Sigma , \cdots
=1,\cdots ,6$. The gauge vectors 
$A^\alpha_{\mu\Delta}$ (with $\alpha =1,2$) come from the N-S and R-R two forms with one internal index.
In terms of the $B^{\Lambda\Sigma}$ the covariant derivatives are
\be
{D}_\mu B^{\Lambda\Sigma}=\partial_\mu B^{\Lambda\Sigma} - f^{\Lambda\Sigma\Delta}_\alpha 
A^\alpha_{\mu\Delta}\ee
where $f^{\Lambda\Sigma\Delta}_\alpha $ are the real N-S and R-R three-form fluxes.
In presence of the fluxes $f^{\Lambda\Sigma\Delta}_\alpha$, the theory develops a scalar potential. In order for 
the potential to have a minimum with vanishing cosmological constant, it is required that the fluxes are
 subject to the constraint
\be f^{-\Lambda\Sigma\Delta}_1 = {\rm i} \alpha f^{-\Lambda\Sigma\Delta}_2 \label{constr}\ee
where $\alpha$ is a complex constant (the minus apex denotes the anti self-dual projection in the internal manifold).
Condition (\ref{constr}) breaks the symmetry $SL(2,\IR) \times GL(6)\subset SL(2,\IR) \times SO(6,6) $ to $U(4)$.

In the vacuum, the  $SL(2,\IR)/U(1)$ variable $S= C+{\rm i} e^{\phi}$  is actually fixed at the value $S=-L^2 /L^1 ={\rm i}\alpha$,
that is, in terms of the N-S and R-R IIB dilatons $(\phi,C)$ 
\be e^{\phi}= \Re \alpha\,, \quad C =-\Im \alpha \ \qquad (\Re \alpha > 0), \label{dil}\ee 
 where $L^\alpha$
is the complex component of the  coset representative satisfying the $SL(2,\IR)$ condition 
$L^\alpha \bar L^\beta - \bar L^\alpha L^\beta = {\rm i} \epsilon^{\alpha\beta}$.
Equation (\ref{dil}) also implies that the N-S and R-R fluxes must both be present to have dilaton stabilization, 
so that the solution is non perturbative at the string level.

The basic features of the model are encoded in the two matrices
\be F^{IJK} \equiv L^\alpha f_\alpha^{IJK} \, ,\qquad \bar F^{IJK} \equiv \bar L^\alpha f_\alpha^{IJK} \ee
where $f_\alpha^{IJK} \equiv f_\alpha^{\Lambda\Sigma\Delta}E^I_\Lambda E^J_\Sigma E^K_\Delta$
and $E^I_\Lambda$ are the coset representatives of $GL(6)/SO(6) \subset SO(6,6)/[SO(6)\times SO(6)]$. 
Indeed, by using the $SU(4) \sim$ spin($SO(6)$) $\Gamma$-matrices one can define the two complex symmetric matrices
\bea
S_{AB} &=& -\frac{\rm i}{48} \bar F^{-IJK}(\Gamma_{IJK})_{AB}\nonumber\\
N^{AB} &=&\frac{1}{48} \bar F^{+IJK}(\Gamma_{IJK})^{AB}
\eea
where $S_{AB}$ is the gravitino mass matrix and $N^{AB}$ the fermionic shift in the dilatino supersymmetry transformation law.
They have the same $U(1)$ charge, but contra-gradient $SU(4)$ representations $\mathbf{ 10}, \mathbf{ \overline{10}}$ respectively.
The scalar potential is actually given in terms of $N^{AB}$:
\be V \propto |N^{AB}|^2\ee

For generic values of the fluxes (still constrained by equation (\ref{constr}))  equation $N^{AB}=0$ stabilizes the dilaton $e^\phi$ and axion $C$ together, while fixing 
$E^I_\Lambda$ to the diagonal form diag$(e^{\phi_1},e^{\phi_2},e^{\phi_3},e^{\phi_1},e^{\phi_2},e^{\phi_3})$.
There is no residual supersymmetry in this case.

Some residual supersymmetry is preserved when some extra constraints among the fluxes are satisfied.
By adopting complex coordinates $I \to (i,\bar \imath)$, $i, \bar \imath =1,2,3$ (so that $F^{IJK} \to 
(f^{ijk},f^{ij\bar k}, f^{i\bar \jmath \bar k}, f^{\bar \imath\bar \jmath \bar k})$), the four gravitino masses are given by
\be 
|f_{ijk}|\,, \quad |f_{i\bar \jmath \bar k}| \,; \qquad \ i\neq j \neq k 
\ee
There are in fact four different values of this type. When any one of them  vanishes, then one supersymmetry remains unbroken.
In this case some of the $E^I_\Lambda$ moduli remain unfixed.

For instance,  for preserving $N=3$ supersymmetry it is needed that $f_{i\bar \jmath \bar k} =0$, while only
$f_{ijk}=f \epsilon_{ijk}\neq 0 $. Then the $g_{ij}$ entries of the scalar metric are fixed to zero, while the $g_{i\bar \jmath}$ 
remain unfixed.
Correspondingly, the six axions $B^{ij}$ are eaten by six of the vectors, while the remaining nine axions $B^{i\bar \jmath}$
 remain massless.
The resulting $N=3$ moduli space is $U(3,3)/[U(3) \times U(3)]$, as predicted by $N=3$ supergravity.
There is a single massive spin 3/2 multiplet, which is long (not BPS saturated) and this is in agreement with the fact that there are not
gauged central charges in this model. This is to be contrasted to what happen with the \Ss $N=8$ spontaneously broken supergravity, where
 all the massive multiplets are 1/2 BPS saturated.

\subsection{The $K3\times T_2 /Z_2$ orientifold}
Let us now turn to discuss the $K3\times T_2 /Z_2$ orientifold model, which has been studied in the literature both
as a compactification from ten dimensional IIB theory as well as a four dimensional gauged supergravity.

From a four dimensional point of view it is   an $N=2$ supergravity model coupled to 3 vector multiplets and 20 hypermultiplets.
 The scalars of the vector multiplets parametrize the special K\"ahler 
manifold\footnote{We only give here the local spectrum of the scalar manifolds,
 avoiding the discussion of discrete identifications.}
\be \cM_V  =\left[\frac{SL(2,\IR )}{SO(2)}\right]^3
= \frac{SL(2,\IR )}{SO(2)}\times  \frac{SO(2,2 )}{SO(2)\times SO(2)}
\ee
while the scalars in the hypermultiplets parametrize the quaternionic manifold
\be 
 \cM_H =\frac{SO(4,20)}{SO(4)\times SO(20)}.\ee
The latter can be regarded as a fibration over 
$\frac{SO(3,19)}{SO(3)\times SO(19)}$
\be \frac{SO(4,20)}{SO(4)\times SO(20)}
=\frac{SO(3,19)}{SO(3)\times SO(19)}\times \IR^++   {\bf 22}\ee
where  the presence on $\cM_H$ of 22 translational isometries $(C^m,C^a)$, ($m=1,2,3$; $a=1,\cdots , 19$) 
corresponding to the degrees of freedom of the 10 dimensional R-R 4-form, with two indices on the $K3$  and two on the torus, is put in evidence.

Switching on three-form fluxes  
corresponds to  gauging  some of the 22  translational isometries
 of $\cM_H$ by some of the vectors $A_\mu^\Lambda$ ($\Lambda =0,1,2,3$)  in the theory, where $\Lambda = (i,\alpha)$, $i \in SL(2,\IR)_{T_2}$, $\alpha \in SL(2,\IR)_{IIB}$.

The covariant derivatives are
\begin{eqnarray} {D}_\mu C_m &=&\partial_\mu
C^m+f^m_{\Lambda}A^{\Lambda}_\mu,\\ {D}_\mu C_a&=&\partial_\mu
C^a+h^a_{\Lambda}A^{\Lambda}_\mu . \end{eqnarray}
Here, the couplings $f^m_{\Lambda}, h^a_{\Lambda}$ correspond to the N-S and R-R three-form fluxes with one index on the torus 
and two on $K3$. 

The presence of fluxes 
allows step-wise partial super-Higgs $N=2 \to N=1 \to 0 $ with 
zero vacuum energy.

 $N=2$   and   $N=1, 0$  vacua correspond
to two gaugings different in the choice of
 quaternionic isometries  and of  gauged vectors. 
In all cases many of the moduli are stabilized.

To obtain configurations with $N=2$  supersymmetry one should switch on the fluxes $h^1_{2}=g_2, h^2_{3}=g_3$, 
 taking as   would be Goldstone bosons the scalars $C^{a=1} , C^{a=2}$.
In this case the vectors which become massive are the vector partner of IIB dilaton and of
$T^2$ complex structure moduli.
In fact,  two of the original massless hypermultiplets and the
two  vector multiplets of $A_2$ and $A_3$ combine into two long
massive vector multiplets $[1,4(\frac 1 2), 5(0)]$.
We see that the $N=2$ configurations are just an example of the
Higgs phenomenon of two vector multiplets. The residual moduli
space is
\be\frac{SO(4,18)}{SO(4)\times SO(18)}\times
\frac{SU(1,1)}{U(1)}.\ee
Let us note that to have $N=2$ preserving vacua it is not
possible to gauge more than two vectors, since it would be incompatible with
 the given symplectic frame, used to reproduce the orientifold configuration.  It is
easy to see that the given choice is the only one stabilizing the
moduli in a way compatible with the domain of definition of the scalar fields. This same result is derived in Section 5
of \cite{tt}, with topological arguments.

\bigskip

 On the contrary, to obtain  configurations with   $N=1,0$ supersymmetry
one should switch on, at least, the couplings  $f^1_{0}=g_0, f^2_{1}=g_1$
corresponding to the scalar isometries $C^{m=1} , C^{m=2}$.  The massive vectors are in this case the 
graviphoton and vector partner of the $ K3$ volume modulus.
 For general values of the fluxes, the supersymmetry is completely broken; 
the $N=1$ case is obtained by imposing  further 
 $|g_0|=|g_1|.$

The massless spectrum of the $N=1$ reduced
theory is the following. From the 58 scalars of $SO(3,19)/(SO(3)\times
SO(19))\times\IR^+$ there remain 20 scalars parametrizing
$SO(1,19)/ SO(19)\times\IR^+$. From the 22 axions there
remain 20. All together they complete the scalar content of 20 chiral
multiplets. 
The spectrum includes two massless vector multiplets
corresponding to $A^2_\mu$ and $A_\mu^3$ and an extra chiral
multiplet whose scalar field comes from the $N=2$ vector
multiplet sector. There is then  one long massive gravitino multiplet 
$[(3/2), 2(1), (1/2)]$,  containing as vectors
the graviphoton $A^0$ and the vector $A^1$.

Note that this $N=2$ model  allows partial Super Higgs, evading the no-go theorem for $N=2$ supergravity \cite{CGP}.
This is possible because the symplectic frame chosen such as to reproduce the string model given uses a degenerate symplectic section
for  special geometry on $\cM_V$, where no prepotential function $F(X)$ exists. 

More general vacua, preserving $N=1,0$ supersymmetry, can be
obtained by considering an arbitrary vector coupling $g_\Lambda$. 
In this more general case, $g_0$ and $g_1$ gauge two of the isometries $C^m$, while $g_2$ and
$g_3$ gauge two of the isometries $C^a$. The $N=1$ preserving vacua have 18 
left over chiral multiplets.

\section*{Acknowledgments}
This report is based on collaborations with C. Angelantonj, R. D'Auria, 
F. Gargiulo, M. A. Lled\'o and S. Vaul\`a, which we would like to thank. 
M.T. would like to thank H. Samtleben for useful discussions.
M.T. would like to thank the Th. Division of CERN, where part of this work has
 been done, for their kind hospitality.
The work of S.F. has been supported in part by European Community's Human 
Potential Program under contract HPRN-CT-2000-00131 Quantum Space-Time, in 
association with INFN Frascati National Laboratories and by D.O.E. grant 
DE-FG03-91ER40662, Task C. The work of M.T. is supported by a European
 Community Marie Curie Fellowship under contract HPRN-CT-2001-01276.

\end{document}